\documentclass[twocolumn,showpacs,aps,prd,amsmath,amssymb,nofootinbib,nobibnotes,floatfix]{revtex4-1}
\usepackage{hyperref}
\usepackage{amsfonts}
\usepackage{amsmath}
\usepackage{mathrsfs}
\usepackage{epsfig}
\usepackage{graphicx}
\usepackage{url}
\usepackage{hyperref}
\usepackage{float}
\usepackage{color}
\usepackage[utf8]{inputenc} 
\usepackage{graphicx}
\usepackage{epsf}
\usepackage{comment}
\usepackage{slashed}
\usepackage{footmisc}
\usepackage{setspace}

\begin{document}

\title{Black Hole Mergers from Globular Clusters Observable by LISA and LIGO: Results from post-Newtonian Binary-Single Scatterings}

\author{Johan Samsing}
\email{Email: jsamsing@gmail.com}
\affiliation{Department of Astrophysical Sciences, Princeton University, Peyton Hall, 4 Ivy Lane, Princeton, NJ 08544, USA}

\author{Daniel J. D'Orazio}
\affiliation{Department of Astronomy, Harvard University, 60 Garden Street Cambridge, MA 01238, USA}

\author{Abbas Askar}
\affiliation{Nicolaus Copernicus Astronomical Center, Polish Academy of Sciences, ul. Bartycka 18, 00-716 Warsaw, Poland}

\author{Mirek Giersz}
\affiliation{Nicolaus Copernicus Astronomical Center, Polish Academy of Sciences, ul. Bartycka 18, 00-716 Warsaw, Poland}

\begin{abstract}

We study the gravitational wave (GW) frequency and chirp mass distribution of binary black hole (BBH) mergers assembled through three-body interactions in globular clusters (GCs),
when GW emission at the 2.5 post-Newtonian (PN) level is included in the $N$-body equation-of-motion (EOM). From performing $\sim 2.5\times10^{6}$ PN binary-single interactions
based on GC data from the `MOCCA-Survey Database I' project, and by the use of analytical methods, we find that $5-10\%$ of all the three-body assembled GC BBH mergers have a GW frequency
at formation that is $\gtrsim 10^{-1}$ Hz, implying they enter the LIGO band without having drifted through the LISA band first. If PN terms are not included in the EOM, one finds
instead that all BBH mergers drifts through both LISA and LIGO. As the fraction of BBH mergers that only show up in LIGO is expected to be $\sim 0\%$ for standard field binary BBH
mergers, future joint measurements with LISA and LIGO can be used to gain insight into the formation of BBH mergers.

\end{abstract}

\maketitle

\section{Introduction}

Gravitational waves (GWs) emitted from the coalescence of binary black holes (BBHs) have been observed \citep{2016PhRvL.116f1102A,
2016PhRvL.116x1103A, 2016PhRvX...6d1015A, 2017PhRvL.118v1101A, 2017PhRvL.119n1101A},
but how and where these BBHs formed are still open questions. 
Several formation channels and environments have been proposed, including dense stellar clusters \citep{2000ApJ...528L..17P, 2010MNRAS.402..371B,
2013MNRAS.435.1358T, 2014MNRAS.440.2714B, 2015PhRvL.115e1101R, 2016PhRvD..93h4029R, 2016ApJ...824L...8R, 2016ApJ...824L...8R,
2017MNRAS.464L..36A, 2017MNRAS.469.4665P}, isolated field binaries \citep{2012ApJ...759...52D, 2013ApJ...779...72D, 2015ApJ...806..263D, 2016ApJ...819..108B, 2016Natur.534..512B},
captures between primordial black holes \citep{2016PhRvL.116t1301B,
2016PhRvD..94h4013C, 2016PhRvL.117f1101S, 2016PhRvD..94h3504C},
active galactic nuclei discs \citep{2017ApJ...835..165B,  2017MNRAS.464..946S, 2017arXiv170207818M},
and galactic nuclei \citep{2009MNRAS.395.2127O, 2015MNRAS.448..754H, 2016ApJ...828...77V, 2016ApJ...831..187A, 2017arXiv170609896H},
but observationally distinguishing these possible pathways from each other is
difficult. Recent studies have pointed out that the BH spin orientations are likely to be different between different channels \citep{2016ApJ...832L...2R};
however, several of the already observed BBH mergers show surprisingly very little, or anti-aligned, spin.
Another possibility is to consider BBH eccentricities, which have been shown to be non-negligible for a relatively large fraction of BBH mergers forming in both classical
globular cluster (GC) systems \citep{2014ApJ...784...71S, 2017ApJ...840L..14S, 2017arXiv171107452S, 2017arXiv171204937R, 2018ApJ...853..140S, 2017arXiv171206186S},
as well as in galactic nuclei \citep[\textit{e.g.},][]{2009MNRAS.395.2127O, 2012PhRvD..85l3005K, 2017arXiv171109989G}.
For example, \cite{2017arXiv171107452S} recently showed, using simple analytical arguments,
that $\sim 5\%$ of all BBHs mergers forming in GCs will have an eccentricity $>0.1$ at $10$ Hz, compared to $\sim 0\%$ for field mergers.
This surprisingly high fraction of eccentric mergers originate from GW capture mergers that form during resonating three-body
interactions \citep[\textit{e.g.},][]{2006ApJ...640..156G, 2014ApJ...784...71S}; a population that only can be probed when General Relativistic (GR) effects
are included in the $N$-body equation-of-motion (EOM). This highly motivates the current development of
eccentric GW templates \citep[\textit{e.g.},][]{2017PhRvD..95b4038H, 2017arXiv170510781G, 2018PhRvD..97b4031H}.

Multi-band GW observations provide other interesting possibilities for constraining
the formation of merging BBH systems \citep{2017arXiv171011187M, 2017arXiv170200786A, 2017ogw..book...43C, 2017ApJ...842L...2C}. For example, as pointed
out by \cite{2016PhRvL.116w1102S, Seto:2016}, the first BBH merger observed (GW150914) could have been seen by a GW instrument similar to the proposed
`Laser Interferometer Space Antenna' (LISA) a few years before entering the band of the
`Laser Interferometer Gravitational-Wave Observatory' (LIGO). A LISA type mission would therefore make it possible to `prepare' for LIGO events,
opening the prospect for detailed studies of precursors to GW sources \citep{2016PhRvL.116w1102S, 2017ApJ...842L...2C}. Other possibilities include measuring the
BBH eccentricity distribution in the LISA band, which has been shown to differ between BBH progenitor channels \citep[\textit{e.g.},][]{2016PhRvD..94f4020N, 2017MNRAS.465.4375N}. 
However, despite these encouraging possibilities, we note that no detailed work has been performed so far on
how BBH mergers dynamically formed in stellar clusters distribute when post-Newtonian (PN) terms \citep[\textit{e.g.},][]{2014LRR....17....2B} in the $N$-body EOM are taken into account.
That is, the newly resolved populations of GW mergers forming during resonating few-body interactions \citep{2017arXiv171206186S, 2017arXiv171204937R} and
the possibility for second generation BBH mergers \citep{2017arXiv171204937R}, have not yet been properly discussed in relation to multi-band GW astrophysics. 
We take the first step in this paper.

In this paper we expand upon the work by \cite{2017ApJ...842L...2C}, who studied how BBH mergers are likely to distribute
as a function of their GW frequency and eccentricity, depending on their formation mechanism. They concluded correctly, that
the majority of BBH mergers that form in resonating binary-single interactions are likely to elude the LISA band. However,
no detailed calculations or simulations were performed, only a few orders of magnitude estimates were done based on the work by \cite{2014ApJ...784...71S}.
To improve on their analysis, we present here a study on how BBH mergers assembled in GCs distribute as a function of their GW frequency at formation
when PN terms are included in the $N$-body EOM, using both detailed numerical and analytical methods, together with GC Monte-Carlo (MC) simulations
performed by the \texttt{MOCCA} (MOnte Carlo Cluster simulAtor) code \citep{Giersz2013}.
As described above, and initially pointed out by \cite{2014ApJ...784...71S}, the largest effect from including PN terms, is the formation of BBH
mergers that form during resonating binary-single interactions; a population we refer to in this paper as {\it binary-single GW mergers}.
In agreement with \cite{2017ApJ...842L...2C}, we indeed find that the majority of these binary-single BBH mergers form with a GW frequency
that is above the frequency range of LISA, but within or just below the range of LIGO. By considering the full distribution of GC BBH mergers,
we are further able to estimate that with the PN terms about $5-10\%$ of all BBHs from GCs will not drift through the LISA band before entering the LIGO band (excluding the fraction
that naturally eludes the LISA band because of low signal-to-noise (S/N) caused by their high eccentricity \citep[\textit{e.g.},][]{2017ApJ...842L...2C}), compared to $\approx 0\%$ in the Newtonian case.
As discussed in \cite{2017ApJ...842L...2C}, a population that only appears in the LIGO band is not expected in the standard isolated field binary scenario, suggesting that the
fraction of all BBH mergers forming in clusters can be estimated by simply measuring the fraction that appears only in the LIGO band.
Our results are described in the following sections.

\section{BBH Mergers in LISA/LIGO}

In this section we study how BBH mergers assembled through three-body interactions in GCs distribute as a function of their
GW frequency and chirp mass at their time of formation, when GW emission at the 2.5 PN level is included in the EOM. Using both
numerical (Section \ref{sec:Post-Newtonian $N$-body Scatterings}) and analytical methods (Section \ref{sec:Analytical Estimate}), we find that
$5 \sim 10\%$ of all the BBH mergers observable by LIGO will never appear in the LISA band.
This result complements the recent study by \cite{2017ApJ...842L...2C}, and
naturally opens up for a wealth of new possibilities for constraining how and where
BBH mergers form using multi-band GW detections.

We note that to estimate the actual observable fraction of BBH mergers that will elude the LISA band, but appear in the LIGO band,
one needs to fold in detailed models for both the design of the instruments \citep[\textit{e.g.},][]{2016PhRvL.116w1102S}, as well as the
mass, distance and orbital distributions of the merging BBHs; quantities that unfortunately are poorly constrained at the moment.
To keep our results clear, we therefore only discuss the part of the distribution that is directly shaped by the dynamics, and the inclusion
of PN terms.

\subsection{Post-Newtonian $N$-body Scatterings}\label{sec:Post-Newtonian $N$-body Scatterings}

\begin{figure}
\centering
\includegraphics[width=\columnwidth]{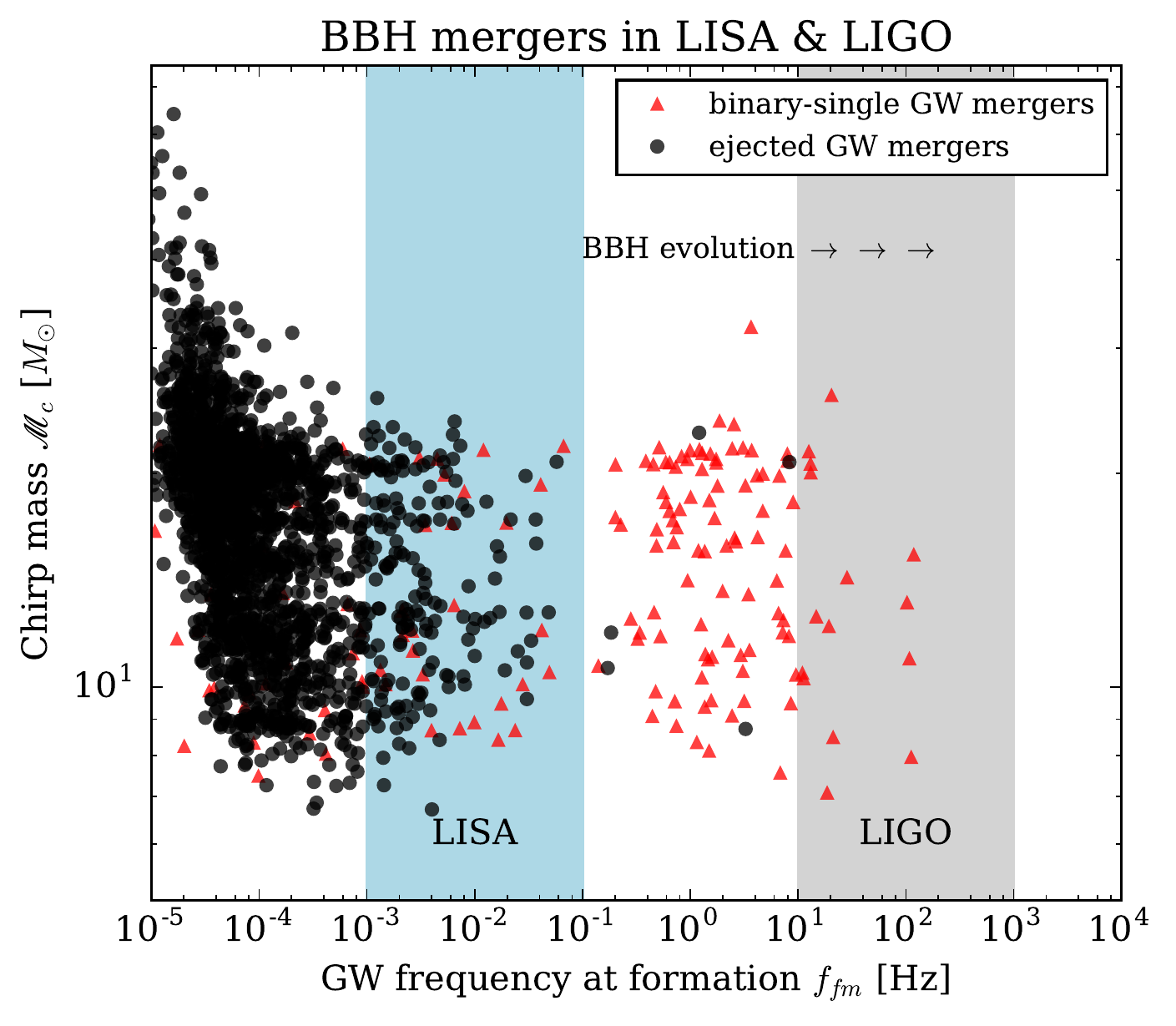}
\caption{Distribution of BBH mergers, as a function of their GW frequency at the time of formation $f_{fm}$ (x-axis) and their source frame chirp mass $\mathscr{M}_{c}$ (y-axis), 
dynamically assembled through binary-single interactions simulated with an $N$-body code that includes GW emission at the 2.5PN
level \citep{2017arXiv171206186S}. As described in Section \ref{sec:Post-Newtonian $N$-body Scatterings}, the initial conditions for the interactions were
extracted from a large set of GC models evolved by the MC cluster code \texttt{MOCCA}, as part of the `MOCCA-Survey Database I' project.
The {\it black points} show the distribution of BBHs that merge outside their host cluster after being dynamically ejected (ejected GW mergers),
where the {\it red points} show the BBHs that merge during a binary-single interaction inside the cluster (binary-single GW mergers).
As seen, the majority of the binary-single GW mergers form with a GW frequency that is above the LISA
band, but within or just below the LIGO band, implying that they will enter the LIGO band without having drifted through the LISA band first (see also \citep{2017ApJ...842L...2C}).
This is in contrast to the classical ejected mergers, that all drift through both the LISA and the LIGO band. As pointed out by \cite{2017ApJ...842L...2C},
measuring the fraction of sources observed in LISA and LIGO can be used to constrain the number of BBH mergers dynamically assembled in clusters.
We note that PN terms are crucial for such a test.}
\label{fig:BBHdist}
\end{figure}

To numerically study the GW frequency distribution of BBHs and the effect from PN terms in the EOM, we used data from  
about $2000$ star cluster models evolved by the \texttt{MOCCA} code \citep[][and references therein]{Giersz2013} as part of the
`MOCCA-Survey Database I' project \citep{2017MNRAS.464L..36A}.
From this set of models, we extracted all the strong binary-single interactions between three BHs each with a mass $<100 M_{\odot}$.
These binary-single interactions were originally evolved in \texttt{MOCCA} simulations using the Newtonian code \texttt{fewbody} \citep{Fregeau2004}.
To study the effect from PN corrections, we therefore re-simulated all of the interactions with our own few-body code that includes the 2.5 PN term
accounting for GW emission \citep[\textit{e.g.},][]{2014LRR....17....2B}. To achieve better statistics, we simulated each interaction $5$ times, which led to a total of $\sim 2.5\times10^{6}$
binary-single simulations. Due to computational limitations, we limited each interaction to a maximum of $2500$ orbital times of the initial target BBH, which led to a $\approx 98\%$ completion fraction.
The results presented below are based on this completed set of PN binary-single interactions. We refer the reader to \cite{2017arXiv171206186S} for a
more detailed explanation of this re-simulation procedure.

The distribution of GW frequencies at formation, $f_{fm}$, and source frame chirp masses, $\mathscr{M}_{c}$, for the binary-single assembled BBH mergers
derived from the \texttt{MOCCA} dataset, as described above, is shown in Figure \ref{fig:BBHdist}. To resolve the distribution evaluated at present time
requires orders of magnitude more simulations than we could perform. The presented distribution therefore includes 
all BBHs that merge after $1$ Gyr and before a Hubble time (By considering the time resolved BBH merger history
derived in \citep{2017arXiv171206186S}, we do expect the distribution shown in Figure \ref{fig:BBHdist} to be similar to the present day distribution).
To derive the GW frequency at formation of a given BBH, $f_{fm}$, we used the approximation presented in \cite{Wen:2003bu}, where we took the time of formation to be
the moment the BBH in question can be treated as an `isolated' binary free from significant perturbations by the unbound (ejected BBH merger) or bound (binary-single GW merger) single BH.
To quantify if a BBH can be treated as `isolated', we used a tidal threshold of $0.1$ as described in \cite{2014ApJ...784...71S}.
As described in the caption of Figure \ref{fig:BBHdist}, the {\it black points} denote the distribution of BBHs that
merge after being dynamically ejected through a binary-single interaction from their GC (ejected BBH mergers), where the {\it red points}
show the BBHs that merge during a binary-single interaction through the emission of GWs (binary-single GW mergers). The red points therefore only appear when
PN terms are included in the EOM.

From considering the distribution shown in Figure \ref{fig:BBHdist}, one concludes that the majority of the BBHs that merge after being dynamically ejected from their GC (black points)
form with $f_{fm} \lesssim10^{-2}$ Hz, implying they will pass through the LISA band before entering the LIGO band. This is in contrast to the 
binary-single GW mergers (red points), which in the majority of cases form with $f_{fm} \gtrsim 10^{-1}$ Hz, and therefore will
enter the LIGO band without having appeared in the LISA band first. From simply counting, we find
that only $\sim 0.1\%$ of the classically ejected BBH mergers will enter the LIGO band without having appeared in the LISA band first,
whereas if one includes the binary-single GW mergers the fraction is instead $5-10\%$. As also noted by \cite{2017ApJ...842L...2C}, the binary-single GW mergers form directly
in the most sensitive region of a detector similar to the proposed DECIGO \citep[\textit{e.g.},][]{2011CQGra..28i4011K, 2018arXiv180206977I}, a result that
undoubtedly will be an interesting science case in addition to those discussed in \cite{2017arXiv171011187M}.

\subsection{Analytical Estimate}\label{sec:Analytical Estimate}

The population of BBH mergers that will appear in the LIGO band without having drifted through the LISA band first,
is greatly dominated by binary-single GW mergers (see Figure \ref{fig:BBHdist}). This trend originates from the fact that
the BBH mergers that form during binary-single interactions must inspiral and merge on a timescale that is comparable to the orbital timescale of the initial target BBH.
This is only possible if the BBH pericenter distance is small \citep[\textit{e.g.},][]{2014ApJ...784...71S}, implying that the corresponding GW frequency will be relatively high.
This is in contrast to the ejected BBH mergers, which are formed with no restriction to their merger time. Because the binary-single GW mergers dominate
the population that enters LIGO and not LISA, we may estimate their fraction analytically, as we will demonstrate below.
For this, we follow the work by \cite{2017arXiv171107452S}, in which the probability for eccentric GW mergers forming in binary-single interactions was
derived for a dense stellar system.

We start by considering two BHs each with mass $m$ in a binary with initial semi-major axis (SMA) $a_{\rm in}$ and eccentricity $e$, that form inside a
dense stellar cluster characterized by an escape velocity $v_{\rm esc}$. As described in \cite{2017arXiv171107452S}, this newly formed BBH will undergo continuous BH
binary-single hardening interactions in the cluster core if $a_{\rm in}$ is less than the local hard binary value \citep[\textit{e.g.},][]{Hut:1983js}.
Each of these interactions lead to an average decrease in the BBH SMA from $a$ to $\delta a$, where the average value of
$\delta$ is $7/9$, assuming the binary energy distribution derived in \cite{Heggie:1975uy}. This hardening process will continue until the interacting
BBH receives a recoil velocity through one of its interactions that is $>v_{\rm esc}$, which is possible if its SMA $a < a_{\rm ej}$, where \citep{2017arXiv171107452S}
\begin{equation}
a_{\rm ej} \approx \frac{1}{6} \left(\frac{1}{\delta} - 1\right) \frac{Gm}{v_{\rm esc}^2},
\end{equation}
after which the BBH is kicked out of the cluster. After leaving the cluster, the BBH will then merge in isolation through the emission
of GW emission. This is the dynamical channel for the classical ejected BBH mergers. However, as described above, GW mergers can also form
during the hardening binary-single interactions before ejection is possible when PN terms are included in the $N$-body EOM \citep{2017arXiv171107452S}.
Considering both of these merger types, one can now argue that the fraction of BBH mergers that only show up in the LIGO band is approximately given
by $P_{\rm bs}/P_{\rm ej}$, where $P_{\rm ej}$ is the probability that an ejected BBH
will merge in isolation within a Hubble time, and $P_{\rm bs}$ is the probability that the BBH instead undergoes a binary-single GW merger during hardening before
ejection is possible \citep{2017arXiv171107452S}. In the following we calculate these two probabilities by integrating over the dynamical hardening history of a typical BBH, assuming the hardening
is dominated by equal mass BH binary-single interactions. We further assume that $P_{\rm bs} \ll 1$, which allows us to treat $P_{\rm ej}$ and $P_{\rm bs}$ as uncorrelated variables.
This is an excellent approximation for classical GC systems, but might break down for dense nuclear star clusters \citep[\textit{e.g.},][]{2016ApJ...831..187A}.

\begin{figure}
\centering
\includegraphics[width=\columnwidth]{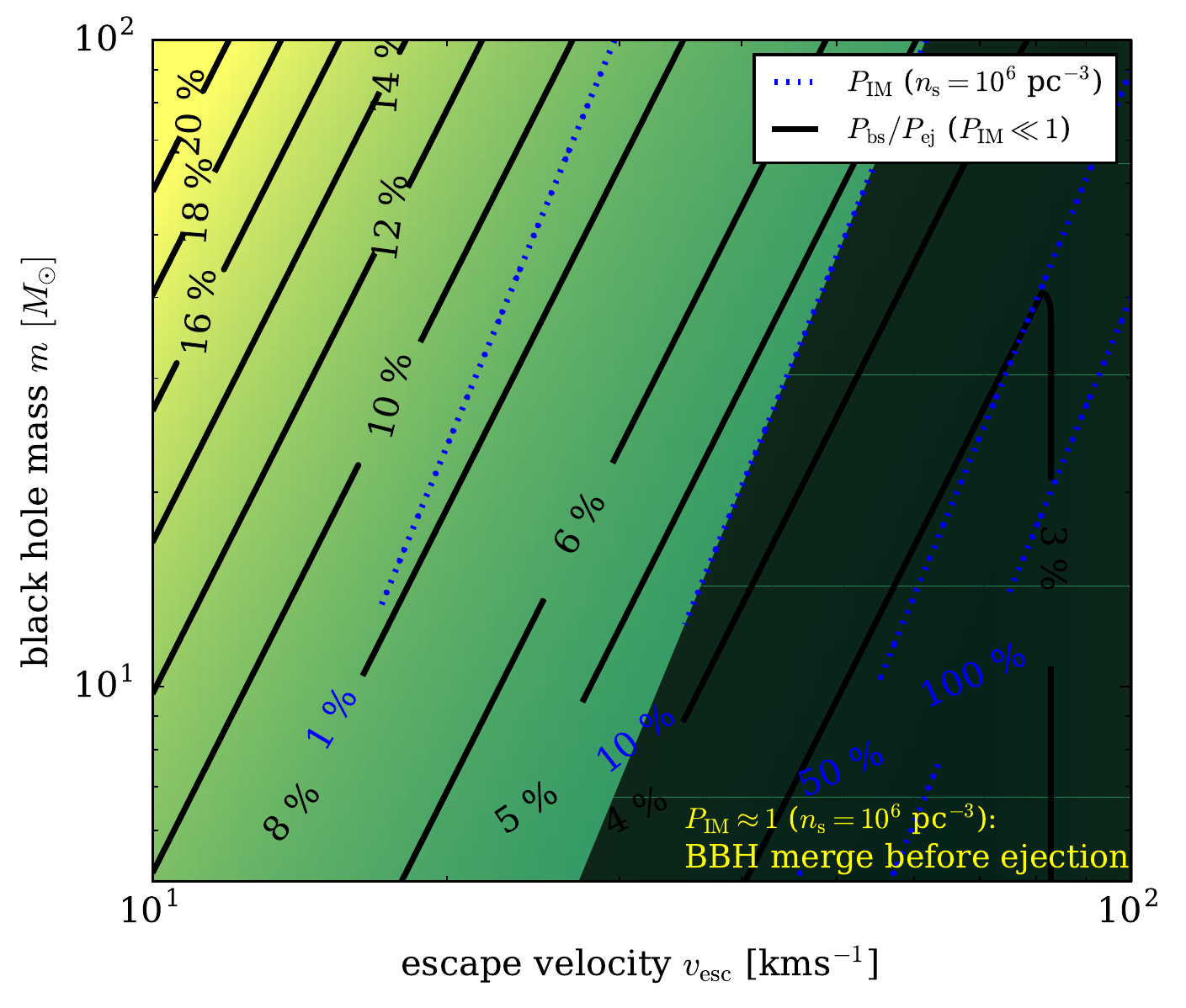}
\caption{The {\it black solid contours} show our analytical estimate of the fraction $P_{\rm bs}/P_{\rm ej}$, where $P_{\rm bs}$ is the probability that a BBH
undergoes a binary-single GW merger during hardening until ejection is possible, and $P_{\rm ej}$ is the probability that, if the BBH is ejected, it will merge
within a Hubble time. This fraction is shown as a function of the host cluster escape velocity $v_{\rm esc}$ (x-axis) and BH mass $m$ (y-axis). As argued
in Section \ref{sec:Analytical Estimate}, the fraction approximately equals the fraction of BBH mergers that will show up in the LIGO band without having drifted through the LISA band.
The {\it blue dotted contours} show the probability that the BBH in question merges in between its binary-single interactions before ejection is possible,
referred to as $P_{\rm IM}$, where IM is short for `Isolated Merger', as further described in \citep{2017arXiv171107452S}. For our analytical estimate
we have assumed that $P_{\rm IM} \ll 1$ and $P_{\rm bs} \ll 1$. The {\it dark shaded region} shows where our analytical approach is believed to break down (assuming a constant
density of single BHs of $n_{\rm s} = 10^{6}$ pc$^{-3}$). As seen, the fraction of BBH mergers that never will appear in the LISA band is $\sim 5-10\%$ for classical GC systems, which is
in good agreement with our numerical results presented in Section \ref{sec:Post-Newtonian $N$-body Scatterings}.
Our analytical approach is described in Section \ref{sec:Analytical Estimate}, which is based on the recent work by \cite{2017arXiv171107452S}.
}
\label{fig:Pfrac}
\end{figure}

The probability that the interaction between a BBH with initial SMA $a$ and a single BH results in a binary-single GW merger during the interaction is, to leading order \citep{2017arXiv171107452S},
\begin{equation}
P_{\rm bs}(a) \approx \frac{2r_{\rm cap}}{a} \times N_{\rm IMS},
\end{equation}
where $N_{\rm IMS}$ is the average number of temporary BBHs formed during the binary-single interaction ($N_{\rm IMS} \approx 20$),
referred to as intermediate state (IMS) binaries \citep{2014ApJ...784...71S}, and $r_{\rm cap}$ is the `characteristic' three-body GW capture pericenter distance. That is, the (maximum) pericenter 
distance two BHs temporarily bound in a resonating three-body state generally need to have for them to undergo a successful GW inspiral merger,
without being disrupted by the bound single BH. The distance $r_{\rm cap}$ changes in principle between each IMS
BBH \citep{2014ApJ...784...71S, 2018ApJ...853..140S}, however, its characteristic value is approximately equal to the
pericenter distance at which the loss of orbital energy through GW emission over one passage is about the total energy of the initial three-body system \citep{2017ApJ...846...36S}. From
comparing the initial three-body energy, which is dominated by the orbital energy of the initial target binary in the hard-binary limit \citep[\textit{e.g.},][]{Hut:1983js}, and the GW energy loss integrated over
one pericenter passage \citep[\textit{e.g.},][]{Hansen:1972il}, follows now that \citep{2017arXiv171107452S}, 
\begin{equation}
r_{\rm cap} \approx {\mathscr{R}_{\rm m}} \times \left({a}/{{\mathscr{R}_{\rm m}}}\right)^{2/7},
\end{equation}
where ${\mathscr{R}_{\rm m}}$ denotes the Schwarzschild radius of a BH with mass $m$.
By now integrating the probability $P_{\rm bs}(a)$ over the series of binary-single interactions that hardens the BBH from $a_{\rm in}$ to $a_{\rm ej}$,
one finds that the total probability that the BBH undergoes a binary-single GW merger before the possibility for ejection, here denoted by $P_{\rm bs}(a_{\rm in}, a_{\rm ej})$,  is given by,
\begin{equation}
P_{\rm bs}(a_{\rm in}, a_{\rm ej}) \approx \int_{a_{\rm ej}}^{a_{\rm in}} \frac{P_{\rm bs}(a)}{a(1-\delta)} da \approx \frac{7}{5}\frac{P_{\rm bs}(a_{\rm ej})}{1-\delta},
\end{equation}
where for the last term we have assumed that $a_{\rm in} \gg a_{\rm ej}$. Note here that we have used that the
differential change in SMA, $da$, per binary-single interaction, $dN_{\rm bs}$, is given by $da = -a(1-\delta)dN_{\rm bs}$.

The probability that the BBH merges within a Hubble time, $t_{\rm H}$, after being ejected from the cluster, ${P}_{\rm ej}(a_{\rm ej})$, 
can be found by combining the GW inspiral life time derived in \cite{Peters:1964bc}, and that the eccentricities of the ejected BBHs tend to
follow a so-called thermal distribution $P(e) = 2e$ \citep{Heggie:1975uy}, as further
described in \citep{2017arXiv171107452S, 2018ApJ...853..140S}. From this it follows that,
\begin{equation}
{P}_{\rm ej}(a_{\rm ej}) \approx
\begin{cases}
(t_{\rm H}/t_{\rm life}(a_{\rm ej}))^{2/7}, & \ t_{\rm life}(a_{\rm ej}) > t_{\rm H}  \\[2ex]
1, & \ t_{\rm life}(a_{\rm ej}) \leq t_{\rm H},
\end{cases}
\end{equation}
where $t_{\rm life}(a_{\rm ej})$ denotes the GW inspiral `circular life time' of the BBH given its SMA $ = a_{\rm ej}$, and its eccentricity $= 0$.

Figure \ref{fig:Pfrac} shows our analytically derived ratio $P_{\rm bs}/P_{\rm ej}$, as a function of $v_{\rm esc}$ and $m$, which to leading order
equates to the expected fraction of BBH mergers that enters the LIGO band without drifting through the LISA band first. The black shaded part shows the region
where our analytical estimate is expected to break down \citep[\textit{e.g.},][]{2017arXiv171107452S}. From the figure, we see that for classical GC systems $P_{\rm bs}/P_{\rm ej} \sim 5-10\%$,
which agrees very well with our full PN scattering results presented in Section \ref{sec:Post-Newtonian $N$-body Scatterings}, although we do note
that our analytical solution is only expected to be valid within a factor of $\sim 2$, due to the assumptions made to make the problem analytically tractable.
However, the apparent excellent agreement is encouraging, and future work will extend our formalism into the nuclear star cluster region.
One also notices that the ratio $P_{\rm bs}/P_{\rm ej}$ is notably dependent on the BH mass $m$, where a greater mass leads to a higher fraction
of mergers that will form above the LISA band. When a large sample of BBH mergers have been observed in multiple GW bands,
this will undoubtedly serve as a useful piece of information when, e.g., extracting the initial mass function of BHs forming in clusters \citep[\textit{e.g.},][]{Kruijssen:2009, Webb+2017}.
We now discuss our findings and conclude.

\section{Conclusions}\label{sec:Conclusions}

We have studied the GW frequency distribution of BBH mergers assembled through
binary-single interactions in GCs, when GW emission at the 2.5 PN level is
included in the EOM. From performing $\sim 2.5\times10^{6}$ PN binary-single
interactions based on GC data extracted from the `MOCCA-Survey Database I'
project \citep{2017MNRAS.464L..36A}, and by the use of the analytical model
presented in \cite{2017arXiv171107452S}, we have illustrated that $5-10\%$ of
the mergers will never drift through the LISA band before entering the LIGO
band. This fraction in the purely Newtonian case is instead $\approx 0\%$,
which clearly illustrates the need for PN cluster simulations; a field that is
just beginning \citep[\textit{e.g.},][]{2017arXiv171204937R, 2017arXiv171206186S}. As
likewise pointed out by \cite{2017ApJ...842L...2C}, the BBH mergers we find
that elude the LISA band when PN terms are taken into account, originate from
BBHs that merge during binary-single interactions through the emission of GWs
\citep{2014ApJ...784...71S}. We now broadly discuss aspects and implications
of our study.

Firstly, the detection of a BBH population that merges in the LIGO band, but
does not pass through the LISA band, would provide indirect evidence for a
dynamical origin of at least a subset of BBH mergers
\citep[\textit{e.g.},][]{2017ApJ...842L...2C}, including the binary-single BBH mergers
resolved in this paper. Not just the existence of, but the fraction of BBHs
formed outside of the LISA band, first computed here for typical GCs, will
provide stronger evidence for their origin. Because this work predicts the
fraction only of BBH mergers assembled in GCs that will not drift through the
LISA band, comparison of this prediction with a measurement of the observed
fraction with future LISA and LIGO observations will allow a determination of
the relative number of BBHs formed through other channels, including isolated
field binary mergers. Dense stellar systems, such as nuclear star clusters,
with and without super massive BHs, would also leave unique imprints across GW
frequency and BBH eccentricity, but no PN work has been done on such systems
yet, and will therefore be the topic of future studies.

A deci-Hz GW detector operating in the frequency range $~0.1-10$~Hz, such as
DECIGO \citep{2011CQGra..28i4011K, 2018arXiv180206977I} or Tian Qin
\citep{TianQin}, would be able to detect GWs in the sensitivity gap between
LISA and LIGO, and thereby probe the formation of not only the dynamically
assembled BBH GW sources studied in this work, but other dynamical channels as
well \citep{2017ApJ...842L...2C}. However, as clearly showed in our work,
precise predictions for their distribution requires a PN treatment.

The orbital parameter distribution of BBHs at formation may also have
implications for the relative GW background in the LISA and LIGO detectors
\citep[\textit{e.g.},][]{ThraneRomano:2013, GW150914_GWB:2016, 2016PhRvL.116w1102S,
Cholis:2017}. Any formation scenario that forms BBHs in the $0.1-10$~Hz band
(or with very high eccentricities at lower frequencies \citep{2017ApJ...842L...2C})
will decrease the level of the high-frequency LISA background that would
otherwise be predicted from assuming that all LIGO mergers pass through the
LISA band \citep[see, \textit{e.g.},][]{2016PhRvL.116w1102S}. We further note that in
addition to the highly eccentric and dynamically assembled systems, there is
another possible BBH formation channel that LISA would miss. For example, the
models of \cite{Loeb:2016, 2017arXiv170604211D} posit the formation of BBHs at
a close separation within the core of a collapsing, hypermassive, rapidly
rotating star. This mechanism may be differentiated from other LISA eluding
scenarios by a preference for high mass, equal mass-ratio binaries and the
prospect of an EM counterpart.

Finally, recent work by \cite{2017ApJ...842L...2C} has shown that highly eccentric BBHs will
also elude detection by LISA. This is because GW emission at the second
harmonic of the orbital period is spread out to higher frequencies, highly
suppressing GW emission until the binary circularizes at closer separations,
in the LIGO-LISA gap. The disentanglement of a highly eccentric vs. a 
GW-capture population, as the one resolved in this work, must be carried out in
future work that tracks the binary eccentricity distribution of the different formation scenarios.
Because a DECIGO-like instrument could directly probe the
region where circularization and GW-capture formation occur, future work
should contrast GW signatures of the two in the $0.1-10$~Hz band.
We do note that if the eccentricity distribution of the classical
ejected BBH merger population is assumed thermal \citep{Heggie:1975uy}, one would be able to
`easily' predict the ratio between BBHs that elude the LISA band due to their eccentricity
and BBHs that simply form at higher frequencies. This suggests that to leading order
source separation is possible. We will follow up on this in future work.

\acknowledgments
JS acknowledges support from the Lyman Spitzer Fellowship.
JS thanks the Niels Bohr Institute, the Kavli Foundation and the DNRF for
supporting the 2017 Kavli Summer Program, and the Nicolaus Copernicus Astronomical Center, Polish Academy of Sciences.
JS and AA also thank the Flatiron Institute's Center for Computational Astrophysics for their generous support during the CCA Numerical Scattering Workshop.
DJD acknowledges financial
support from NASA through Einstein Postdoctoral Fellowship award
number PF6-170151.
AA and MG were partially supported by the Polish National Science Center (NCN), Poland, through the grant UMO-
2016/23/B/ST9/02732. AA is also supported by NCN through the grant UMO-2015/17/N/ST9/02573.

\bibliographystyle{h-physrev}
\bibliography{NbodyTides_papers}

\end{document}